\documentclass[twocolumn,natbib,final]{svjour3}

\usepackage{graphicx}
\usepackage{subfigure,caption}
\usepackage{amsmath}
\usepackage{amssymb,xcolor}
\usepackage{lineno}
\usepackage{ifthen}
\usepackage[noprefix]{nomencl}
\usepackage{ifpdf}
\ifpdf 
\usepackage[linktocpage,pdftex,hyperfigures]{hyperref}
\makeatletter
\providecommand{\doi}[1]{%
  \begingroup
    \let\bibinfo\@secondoftwo
    \urlstyle{rm}%
    \href{http://dx.doi.org/#1}{%
      doi:\discretionary{}{}{}%
      \nolinkurl{#1}%
    }%
  \endgroup
}
\makeatother
\else
\makeatletter
\usepackage{url}
\providecommand{\doi}[1]{%
  \begingroup
    \let\bibinfo\@secondoftwo%
    \urlstyle{rm}{doi:#1\discretionary{}{}{}}%
  \endgroup
}
\makeatother
\fi
\journalname{Microgravity Science and Technology}

\renewcommand{\nomgroup}[1]{\medskip
\ifthenelse{\equal{#1}{B}}{\item[\textit{Abbreviations}]}{ \ifthenelse{\equal{#1}{G}}{\item[\textit{Greek
symbols}]}{ \ifthenelse{\equal{#1}{S}}{\item[\textit{Subscripts}]}{
\ifthenelse{\equal{#1}{P}}{\item[\textit{Superscripts}]}{}}}}}

\makeatletter
\renewcommand{\thesubfigure}{\alph{subfigure}}
\renewcommand{\@thesubfigure}{(\thesubfigure)\hskip\subfiglabelskip}
\makeatother

\begin{document}
\title{Pulsating heat pipe simulations: impact of PHP orientation}
\author{Iaroslav Nekrashevych\and Vadim S. Nikolayev}
\institute{Service de Physique de l'Etat Condens\'e,\\ CEA, CNRS, Universit\'e Paris-Saclay, CEA Saclay,\\ 91191 Gif-sur-Yvette Cedex, France\\
               \email{vadim.nikolayev@cea.fr}}
\authorrunning{I. Nekrashevych\and V. Nikolayev} 

\date{Received: 28 September 2018 / Accepted: 5 February 2019}

\maketitle

\begin{abstract}
The pulsating (called also oscillating) heat pipe (PHP) is a simple capillary tube bent in meander and filled with a two-phase fluid. We discuss numerical simulations of the 10-turn copper-water PHP under vertical favorable (bottom-heated), vertical unfavorable (top-heated), and horizontal orientations. Within the present approach, the horizontal orientation is equivalent to the microgravity conditions. The simulations are performed with the in-house CASCO software. The time-averaged spatial distribution of the liquid plugs inside the PHP is influenced by gravity. This affects the overall PHP performance. We show that, independently of the PHP orientation, contribution of the latent heat transfer is large with respect to the sensible heat transfer. We discuss the phenomena occurring inside the PHP during startup and the stable regimes (intermittent and continuous oscillations followed by dryout).
\keywords{Pulsating heat pipe\and Oscillation\and Liquid films\and Phase change \and Simulation}\end{abstract}

\begin{thenomenclature}

 \nomgroup{A}

  \item [{$d$}]\begingroup tube inner diameter (m)\nomeqref {0}
		\nompageref{1}
  \item [{$F$}]\begingroup evaporator coverage fraction by a phase\nomeqref {0}
		\nompageref{1}
  \item [{$g$}]\begingroup gravity acceleration (m$^2$/s)\nomeqref {0}
		\nompageref{1}
  \item [{$L$}]\begingroup length (m)\nomeqref {0}\nompageref{1}
  \item [{$P, \dot Q$}]\begingroup power (W)\nomeqref {0}\nompageref{1}
  \item [{$R_{th}$}]\begingroup heat transfer resistance (K/W)\nomeqref {0}
		\nompageref{1}
  \item [{$T$}]\begingroup temperature (K)\nomeqref {0}\nompageref{1}
  \item [{$t$}]\begingroup time (s)\nomeqref {0}\nompageref{1}

 \nomgroup{G}

  \item [{$\phi$}]\begingroup volume fraction of liquid in PHP\nomeqref {0}
		\nompageref{1}

 \nomgroup{P}

  \item [{$lat$}]\begingroup latent\nomeqref {0}\nompageref{1}
  \item [{$sens$}]\begingroup sensible\nomeqref {0}\nompageref{1}

 \nomgroup{S}

  \item [{$a$}]\begingroup adiabatic\nomeqref {0}\nompageref{1}
  \item [{$c$}]\begingroup condenser\nomeqref {0}\nompageref{1}
  \item [{$d$}]\begingroup dry\nomeqref {0}\nompageref{1}
  \item [{$e$}]\begingroup evaporator\nomeqref {0}\nompageref{1}
  \item [{$f$}]\begingroup liquid film\nomeqref {0}\nompageref{1}
  \item [{$fb$}]\begingroup feedback section (vertical in Fig.~\ref{PHPviewer})\nomeqref {0}
		\nompageref{1}
  \item [{$l$}]\begingroup liquid plug\nomeqref {0}\nompageref{1}
  \item [{$nucl$}]\begingroup nucleated\nomeqref {0}\nompageref{1}
  \item [{$sat$}]\begingroup at saturation\nomeqref {0}\nompageref{1}
  \item [{$thr$}]\begingroup threshold\nomeqref {0}\nompageref{1}

\end{thenomenclature}

\nomenclature[p]{$sens$}{sensible}
\nomenclature[p]{$lat$}{latent}
\nomenclature[s]{$e$}{evaporator}
\nomenclature[s]{$f$}{liquid film}
\nomenclature[s]{$l$}{liquid plug}
\nomenclature[s]{$a$}{adiabatic}
\nomenclature[s]{$c$}{condenser}
\nomenclature[s]{$d$}{dry}
\nomenclature[s]{$sat$}{at saturation}
\nomenclature[s]{$nucl$}{nucleated}
\nomenclature[s]{$fb$}{feedback section (vertical in Fig.~\ref{PHPviewer})}
\nomenclature[s]{$thr$}{threshold}
\nomenclature[a]{$P, \dot Q$}{power (W)}
\nomenclature[a]{$T$}{temperature (K)}
\nomenclature[a]{$g$}{gravity acceleration (m$^2$/s)}
\nomenclature[a]{$L$}{length (m)}
\nomenclature[a]{$F$}{evaporator coverage fraction by a phase}
\nomenclature[a]{$R_{th}$}{heat transfer resistance (K/W)}
\nomenclature[a]{$t$}{time (s)}
\nomenclature[a]{$d$}{tube inner diameter (m)}
\nomenclature[g]{$\phi$}{volume fraction of liquid in PHP}
\section{Introduction}
\begin{table}[hb]
	\centering
	\caption{Parameters used for the simulation.}
	\begin{tabular}{|l|l|}
		\hline
		Length of one evaporator tube section& $L_e=126$ mm\\
		Length of adiabatic section & $L_a=126$ mm\\
		Length of one condenser tube section & $L_c=110$ mm\\
		Number of turns & $N_{turns}=10$\\
		Feedback section length & $L_{fb}=130$ mm\\
		Filling ratio  & $\phi=0.5$\\
		Inner diameter & $d=1.4$ mm\\
		Outer diameter & 3.2 mm\\
		Condenser temperature & $T_c=22^{\circ}$C\\
		Working fluid  & Water\\
		Tube material  & Copper\\
		Time step & $\Delta t=0.01$ ms\\
		Tube mesh size & 2 mm\\
		Film thickness & $20\;\mu$m\\
		Bubble deletion threshold & $L_{thr}=10\;\mu$m\\
		Nucleated bubble length & $L_{nucl}=100\;\mu$m\\
		Nucleation barrier & $\Delta T_{nucl}=15^{\circ}$C\\
		\hline
	\end{tabular}
	\label{table_prop}
\end{table}
The pulsating (called also oscillating) heat pipe (PHP) is a simple capillary tube bent in meander and filled with a pure fluid that forms inside the vapor bubbles and liquid plugs. When the heat power is applied to one meander (evaporator) side, the sequence of bubbles and plugs inside the tube begins to oscillate spontaneously. This oscillation causes an efficient heat transfer from evaporator to the cooled meander side (condenser). In spite of such a simple structure, the PHP functioning is not completely understood both from experimental and theoretical points of view \citep{EncycExp18}. Tools for the PHP dimensioning are still absent. The numerical simulation seems to be the only way to predict the heat transfer for a given PHP design.

There are many works where the influence of PHP orientation on the PHP performance has been studied experimentally \citep{EncycExp18}. By testing a 10-turn transparent (teflon) PHP filled with R-114 refrigerant during a parabolic flight, \citet{Gu05} found that its performance was better in microgravity that in the vertical favorable position. However most authors agree that the PHP performance in the vertical favorable orientation (evaporator below condenser, called $90^{\circ}$ below) is better than in the horizontal ($0^{\circ}$) orientation, which is in its turn, more advantageous that the vertical unfavorable (evaporator above condenser, $-90^{\circ}$)  orientation. In the present work we simulate a 10-turn water-copper PHP by considering the flow pattern inside the PHP and heat transfer for these three orientations. In other words, we compare the functioning of PHP under the Earth gravity influence (for $90^{\circ}$ and $-90^{\circ}$) and in microgravity (within the present simulation, equivalent to $0^{\circ}$).

The principles of PHP simulation have been established by \citet{shafii1}. Because of the non-stationary functioning that involves multiple vapor liquid interfaces, the 3D or even 2D direct numerical simulation are too costly; 1D modeling appears to be an only viable choice to make the parametric investigation. The simulation techniques has been recently reviewed in detail by \citet{EncycSimu18}. To our knowledge, the orientation effect has been studied by simulation only by \citet{MameliMST12}. However their study used the code of \citet{Holley05} who for some reason used an incorrect vapor energy equation (the initial approach of \citet{shafii1} was entirely correct), see \citet{EncycSimu18} for the discussion.

We present here the simulation results obtained with CASCO (French abbreviation of Code Avanc\'e de Simulation de Caloduc Oscillant:  Advanced PHP Simulation Code, cf. Fig.~\ref{PHPviewer}). It is the 1D simulation in-house software \citep{JHT11}. The plug flow regime (most commonly observed in PHP) can be simulated. Thin liquid films deposited by the receding liquid menisci on the internal tube walls are assumed to be of fixed thickness (cf. Table \ref{table_prop}). This is a ``minimal'' model adopted in many previous simulation approaches (including those cited above). The film thickness value depends on the velocity of the receding meniscus (cf. \citet{IJHMT13} for the discussion). The characteristic meniscus velocity adopted here is 0.7~m/s, which is an average value that we observe in our simulations. The film thickness is kept fixed for all simulation runs presented here to exclude the thickness effect on the results. Unlike older approaches, the liquid films around bubbles have a varying length defined by meniscus motion and film evaporation dynamics \citep{IarATE17}.

Similarly to previous simulation approaches, the gravity is assumed to be constant along each branch; its variation along the curved parts is neglected while the additional pressure drop in the curved parts has been accounted for. In the present simulation, the vapor phase model of \cite{IHPC18VN} is used. More details about the simulation model were described previously by \cite{IarATE17}.

\section{Oscillation startup at different orientations}\label{StartUp}

\begin{figure}[htb]
	\subfigure[]
	{\includegraphics[width=7.6cm,clip]{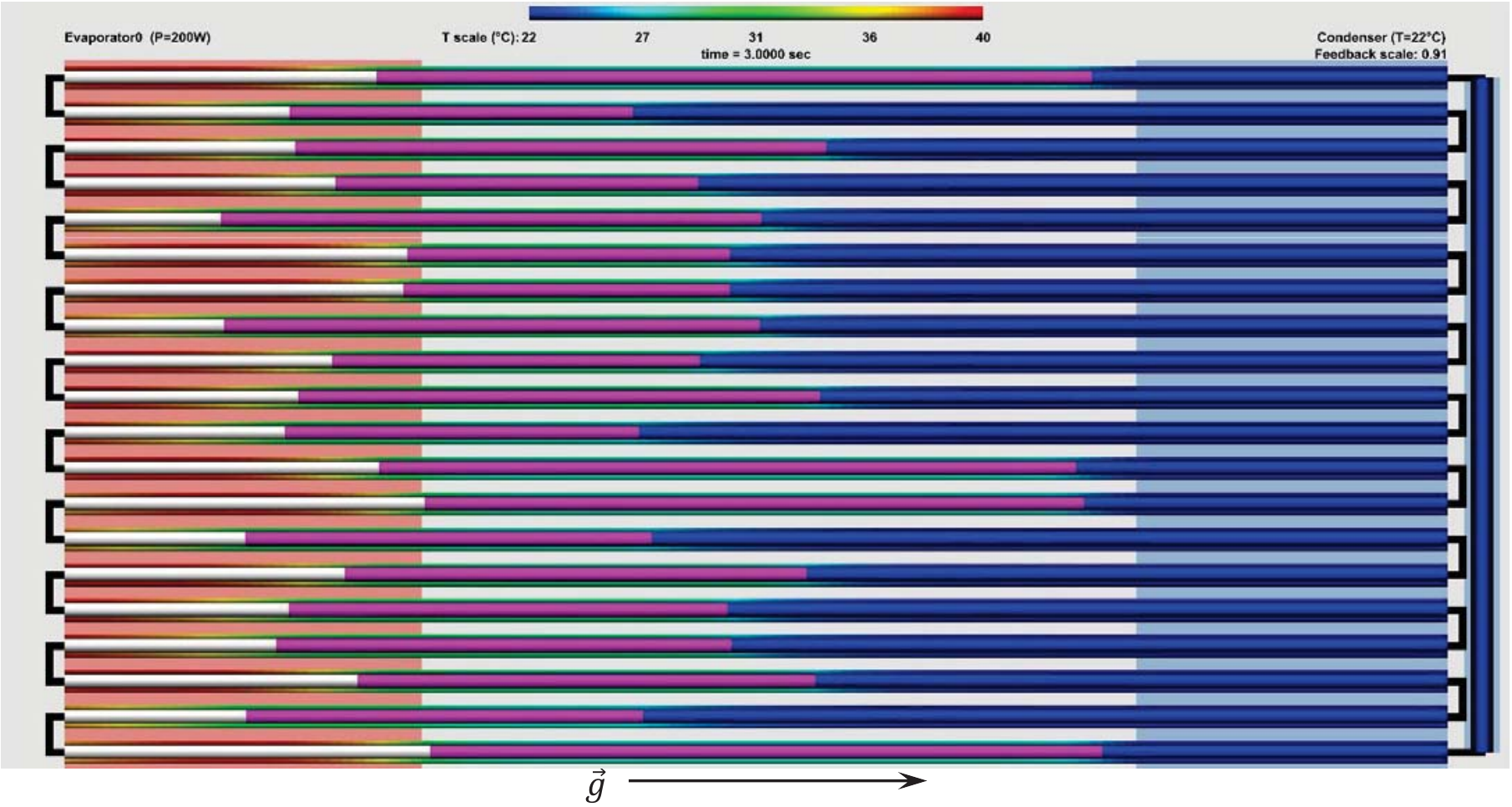}\label{PHPviewer_againstG}}
	\subfigure[]
	{\includegraphics[width=7.8cm,clip]{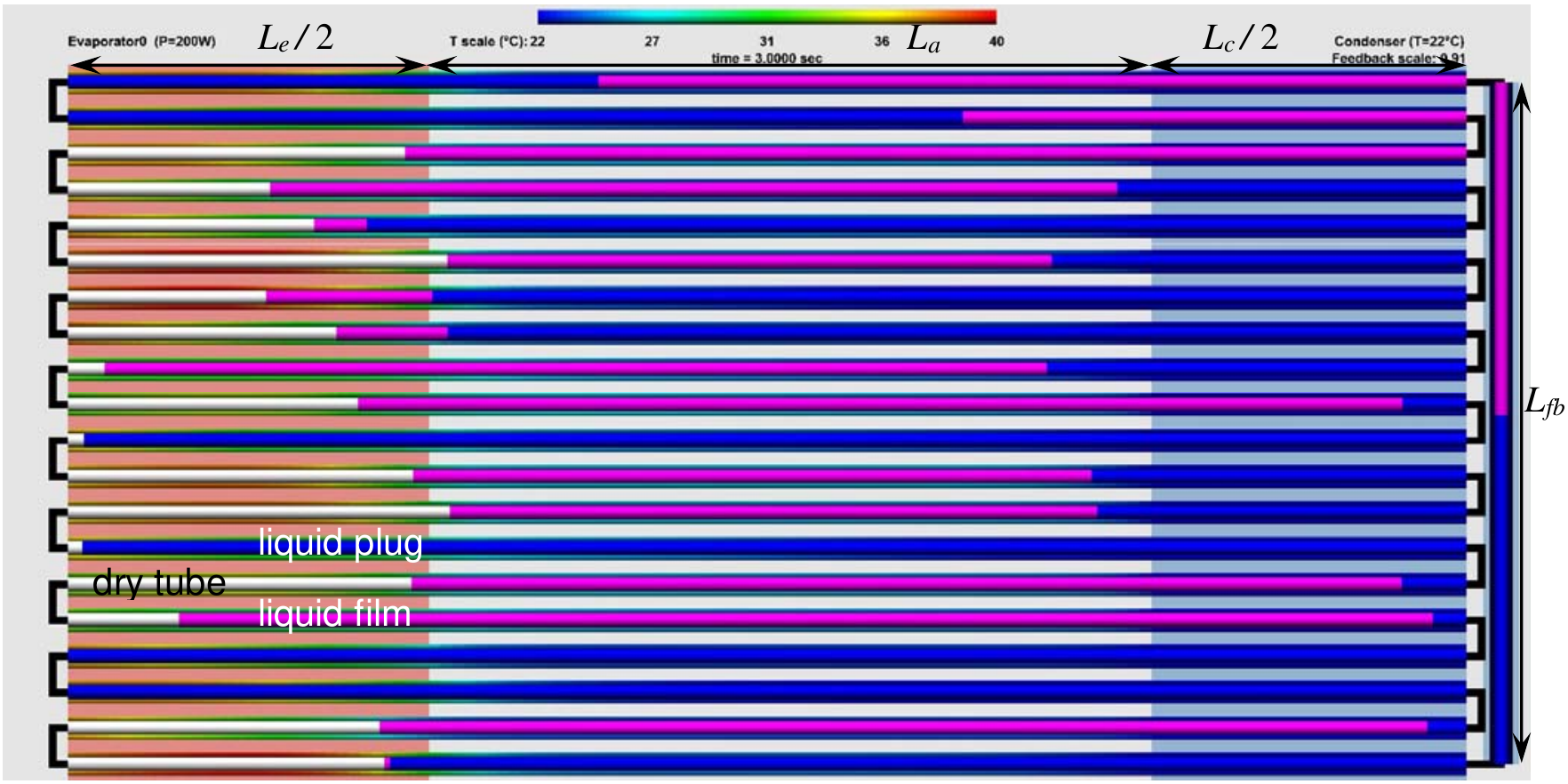}\label{PHPviewer_woG}}
	\subfigure[]
	{\includegraphics[width=7.6cm,clip]{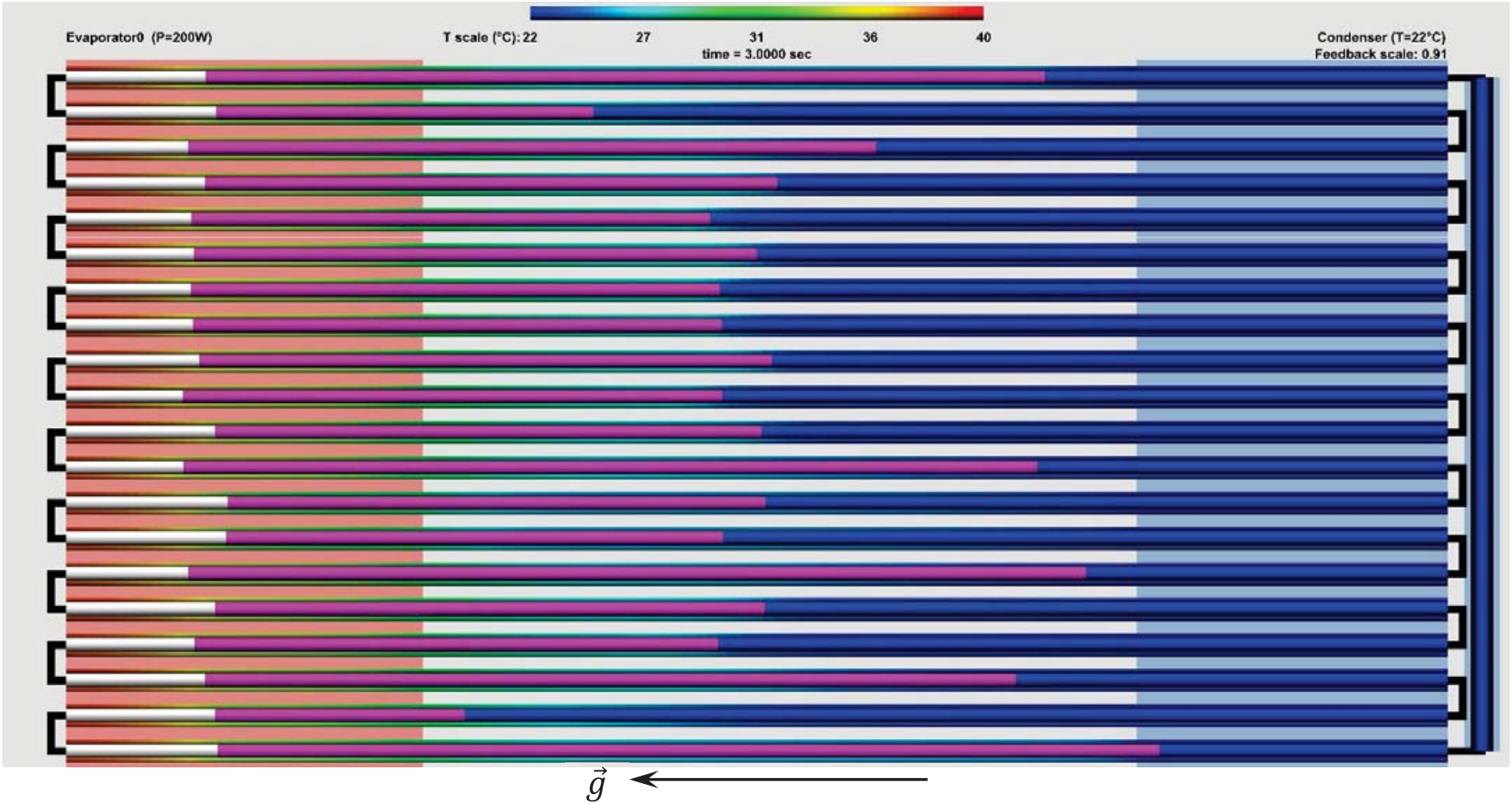}\label{PHPviewer_withG}}
\caption{Start-up phase distribution for $P_e=200$~W and $t=3$~s pictured by the CASCO software for different orientations: (a) $-90^{\circ}$, (b) $0^{\circ}$, and (c) $90^{\circ}$. The direction of gravity is shown by arrow for two vertical orientations. The round turns are not represented for simplicity; black lines are simply connectors linking the equivalent points of neighboring branches. Thin liquid films (in violet) cover the internal tube walls inside the vapor bubbles, except of the dry areas (white), see the image (b). The liquid plugs are blue. The tube wall temperature is shown with color varying from blue to red, which corresponds to $22^{\circ}-40^{\circ}$C interval. The notations for the PHP sizes are shown in the image (b). }\label{PHPviewer}
\end{figure}
The thermal boundary conditions chosen for the simulation are the imposed heat power per unit length of the tube in evaporator (which mimics e.g. the heating with electric isolated wires wound around the tube) and fixed temperature $T_c$ on the internal tube walls in the condenser. The feedback section (vertical in Figs.~\ref{PHPviewer}) is assumed to situate in the middle of the condenser section and inside the condenser block. Since there is no thermal spreader in the evaporator, thermal coupling between the PHP turns is absent there. Such a design is not practical for applications because the dryout in one turn can cause the overall PHP dryout. However this design reveals the local thermal coupling between the tube wall and the fluid. The total evaporator power $P_e$ and other PHP material and working fluid properties used in simulation are shown in Table~\ref{table_prop}.

Initially (at $t=0$ s), the whole PHP is isothermal at $T_c$. Between 10 and 20 liquid plugs of the same length are distributed uniformly throughout the PHP tube. The start-up scenario depends on the PHP orientation. As previously discussed \citep{IarATE17}, because of expansion of the vapor bubbles initially located in the evaporator, all the bubbles outside evaporator are compressed and disappear by condensation. This occurs for all orientations, typically in less than a second after the $P_e$ switching. After this, the plug distribution is the same for different orientations, see Figs.~\ref{PHPviewer_againstG},\ref{PHPviewer_withG}. For this reason, further PHP evolution is almost independent on the initial liquid phase distribution. We checked this by varying the initial number of plugs and adding/removing liquid films between them.

\begin{figure*}[thb]
	\centering
	\subfigure[$-90^{\circ}$ orientation.]
	{\includegraphics[width=5.5cm,clip]{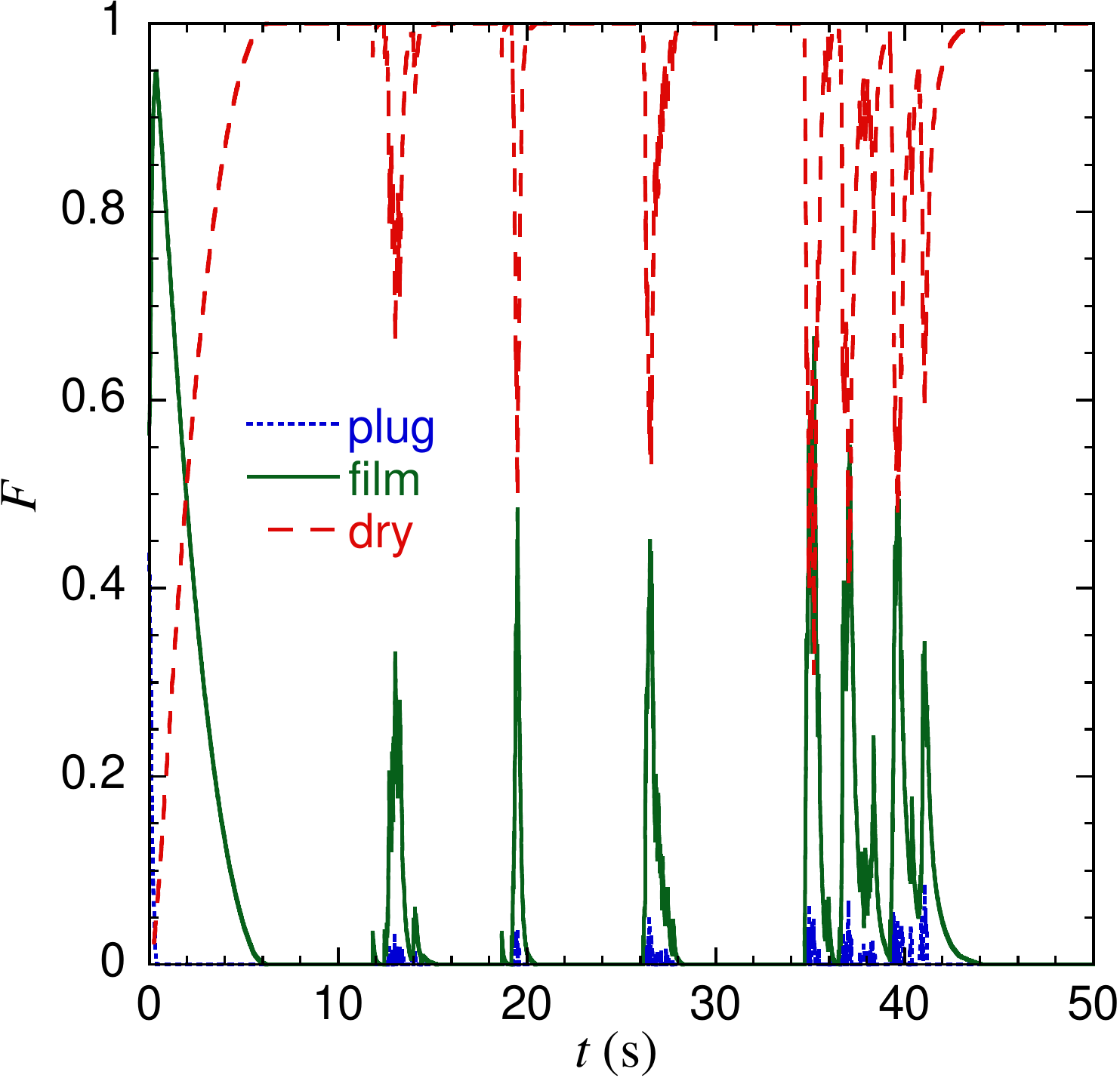}\label{F_start_AG}}
	\subfigure[$0^{\circ}$ orientation.]
	{\includegraphics[width=5.5cm,clip]{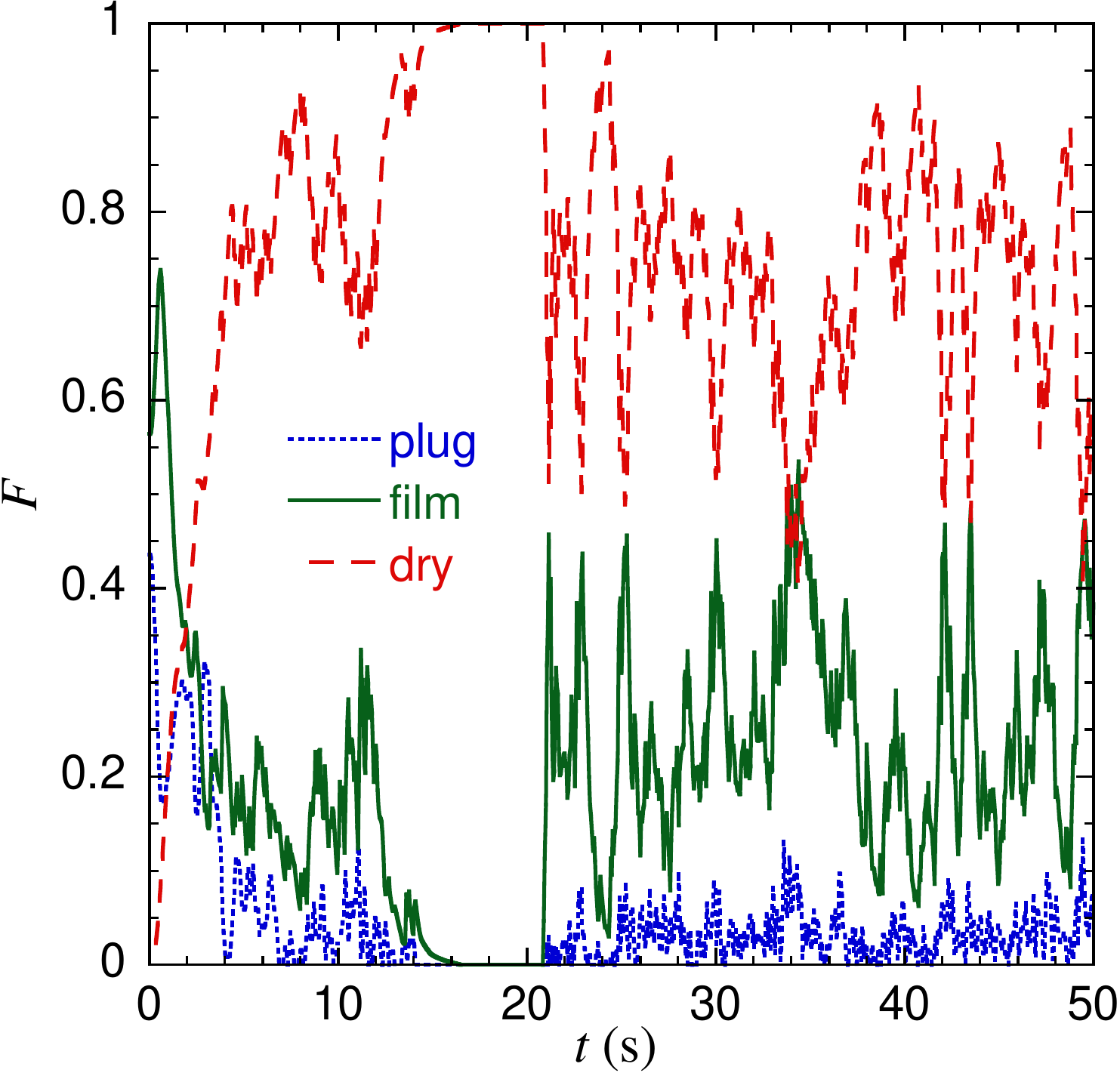}\label{F_start_woG}}
    \subfigure[$90^{\circ}$ orientation.]
	{\includegraphics[width=5.5cm,clip]{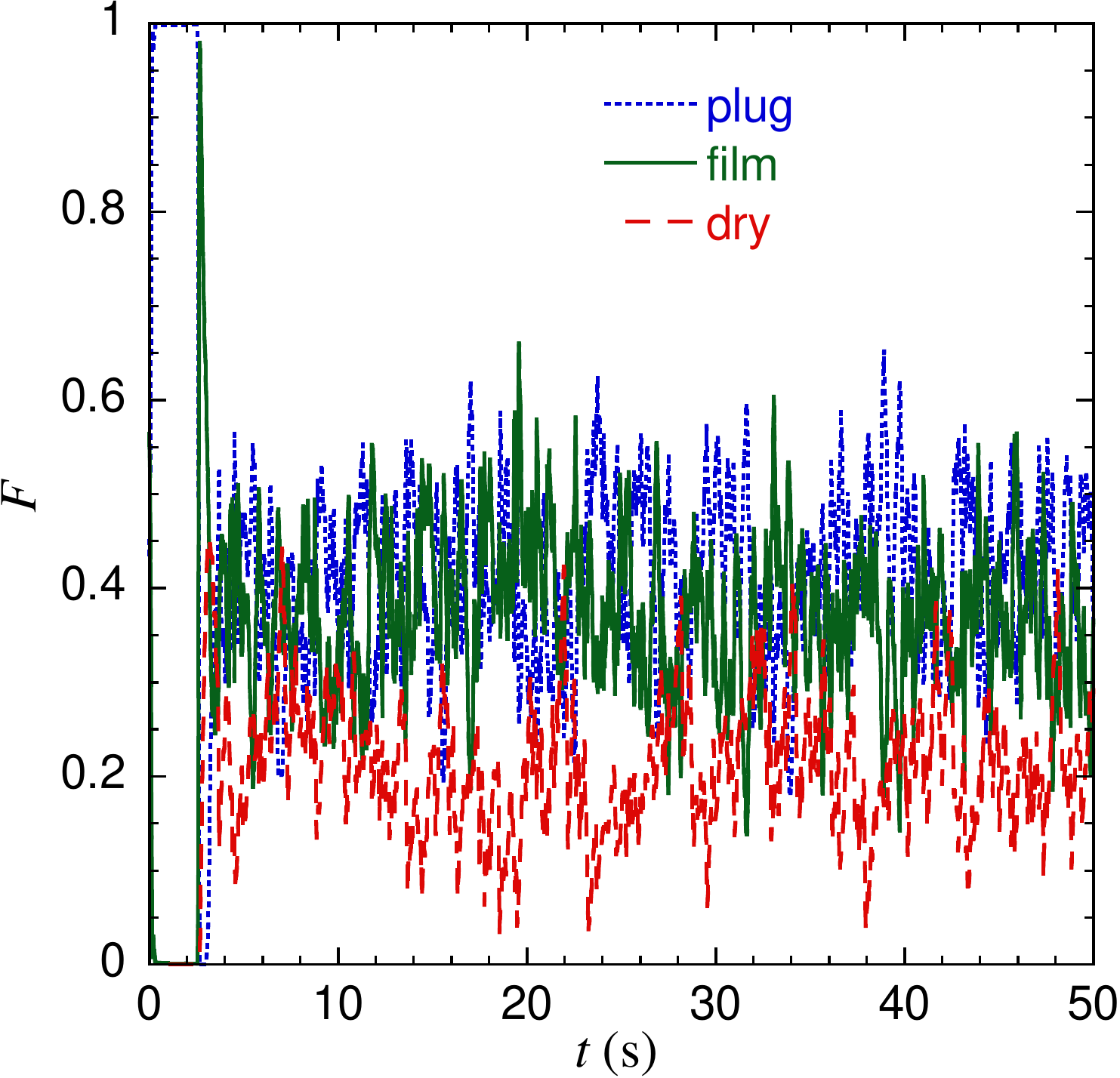}\label{F_start_wG}}
	\caption{Evaporator coverage by liquid plug, liquid film and dry area during the first 50~s of functioning  at different PHP orientations for $P_e=200$~W.}\label{F_start}
\end{figure*}

The distribution of different fluid phases (liquid plugs, liquid films and dry areas) in the evaporator are considered below for each orientation. The distribution of a particular phase can be characterized by an averaged over evaporator length fraction $F$. One distinguishes the fraction covered by the liquid plugs $F_l$, liquid films $F_f$ and dry area $F_d$ (Figs.~\ref{F_start}); at each time moment $F_l+F_f+F_d=1$.

\begin{figure*}[htb]
	\centering
	\subfigure[]
	{\includegraphics[width=5.5cm,clip]{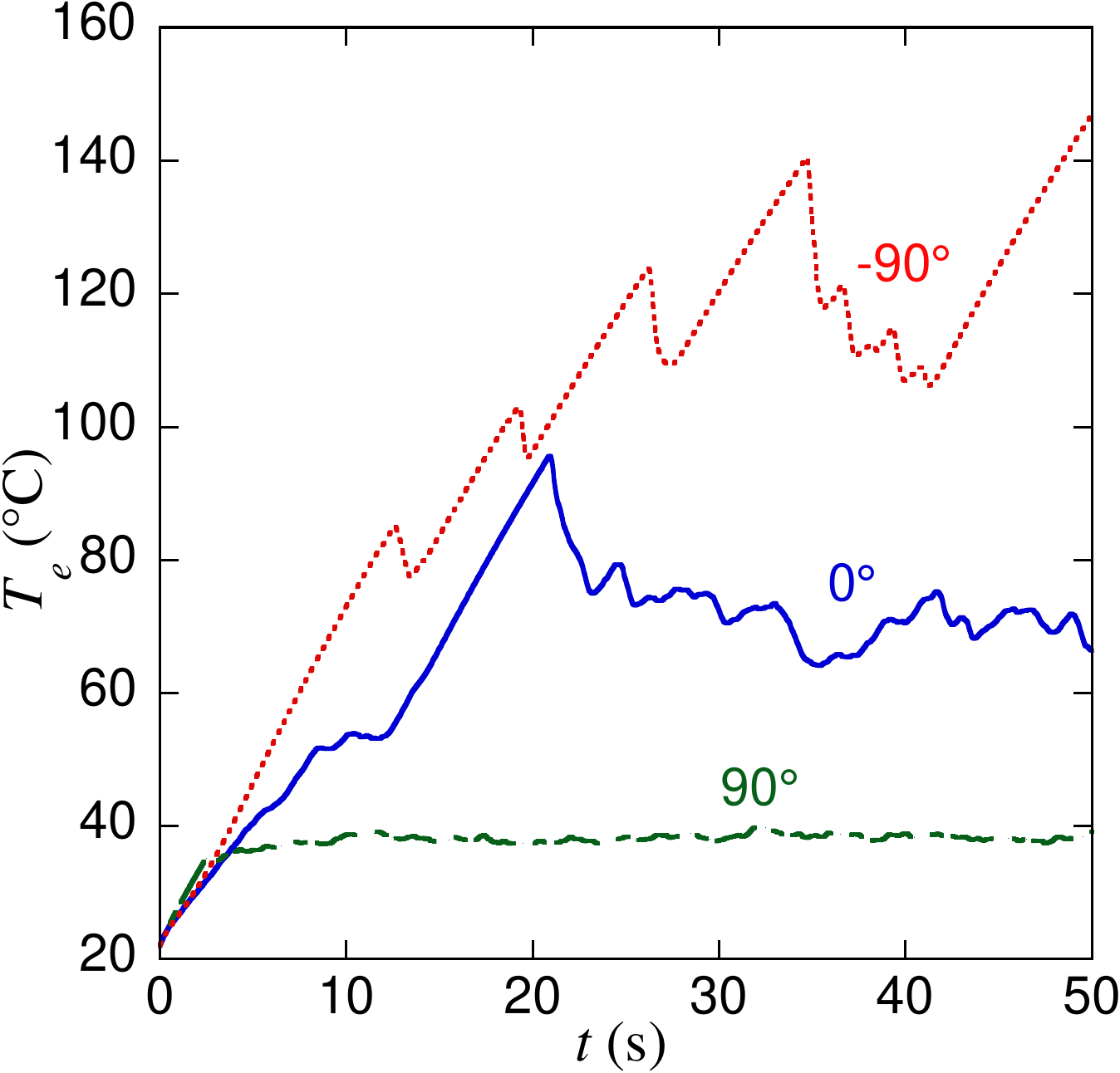}\label{Te_3cases}}
	\subfigure[]
	{\includegraphics[width=5.5cm,clip]{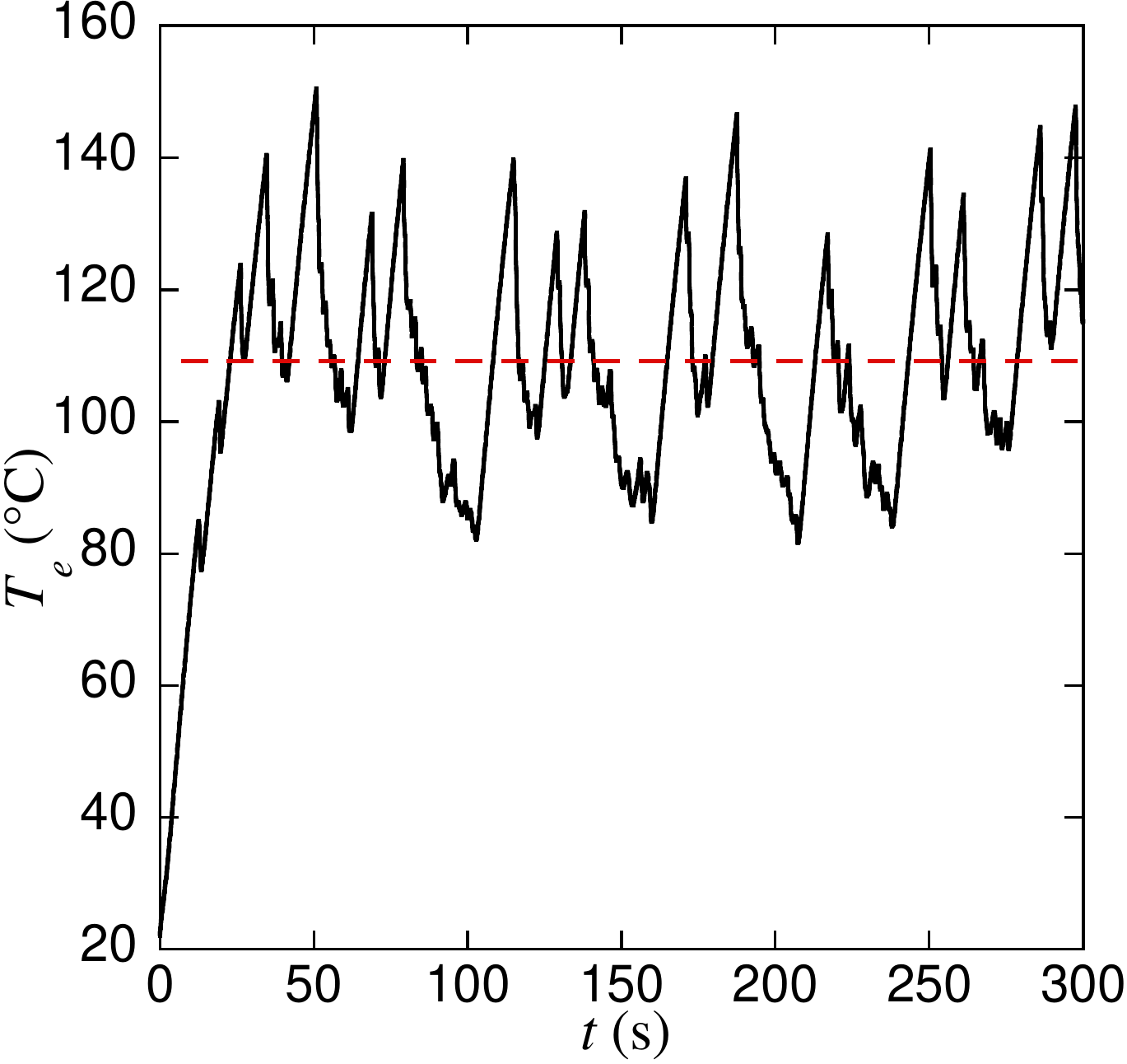}\label{TeAgLong}}
	\caption{Evaporator temperature evolution for $P_e=200$~W. (a) Short-time curves (during the first 50~s of functioning) for different PHP orientations and (b) Long time evolution for $-90^{\circ}$ case. The dashed line shows the average temperature $\langle T_e\rangle\simeq 109^\circ$C for the stable intermittent regime.}\label{TeBoth}
\end{figure*}

In the $-90^{\circ}$ (unfavorable) case, all the liquid gathers in the condenser during the first second as discussed above (Fig.~\ref{PHPviewer_againstG}). The plug fraction $F_l$ drops to zero, see Fig.~\ref{F_start_AG}. The film edges recede ($F_f$ decreases while $F_d$ increases) and the vapor pressure grows due to the film evaporation. The vapor pressure becomes so high that the plugs lose their ability to penetrate into the evaporator. This causes entire evaporator (and partially adiabatic zone) overheating, and the oscillatory motion ceases at about $t=2$~s for the case of $P_e=200$~W (cf. $-90^{\circ}$ curve in Fig.~\ref{Te_3cases}). The averaged over evaporator tube temperature $T_e$ grows. However, this is not a dryout state as the film edges keep receding, which signifies their evaporation; it is just the PHP stopover. At some point (which typically coincides with the disappearance of the film adjacent to the closest to evaporator liquid meniscus), the menisci suddenly begin to oscillate with the increasingly large amplitude (this occurs at $t\simeq 12$~s, cf. $-90^{\circ}$ curve in Fig.~\ref{Te_3cases}) until one of them penetrates into the evaporator so $F_l$ rises. One or several vapor bubbles nucleate and grow inside the fluid. This bubble growth causes a displacement of neighboring liquid plugs to other hot evaporator sections, so bubbles can be nucleated there. As a result, the whole PHP starts-up within a fraction of a second and $T_e$ sharply drops.

For the $90^{\circ}$ (favorable) orientation (Fig.~\ref{PHPviewer_withG}), all the liquid first gathers in the evaporator but the bubbles nucleated there soon expel the liquid into the condenser in spite of the gravity force. The phase distribution becomes the same as in the $-90^{\circ}$ case (Fig.~\ref{PHPviewer_againstG}). Being hydrostatically unstable, this configuration is destabilized much sooner than in the $-90^{\circ}$ case where it is stable. The liquid plugs penetrate into the evaporator at $t\simeq 4$~s for $P_e=200$~W, cf. Fig.~\ref{F_start_wG}. Bubble nucleation begins and the stationary oscillations start-up. Figures~\ref{TeBoth} show that the functioning becomes stable in $90^{\circ}$ case much earlier than in the $-90^{\circ}$ case.

For the $0^{\circ}$  orientation, in the beginning of evolution, some bubbles (among those situating in the evaporator) begin to expand at the expense of the others. The plugs are large so their inertia is strong. The smaller bubbles (even those situating in the evaporator) compressed during the plug motion. They cannot stop the plug motion and finally disappear, i.e. the liquid plugs coalesce. This means that thee plugs become longer; their inertia becomes even larger, which causes further disappearance of small bubbles. As a result, several long liquid plugs (longer than one PHP branch) form during first several seconds so the evaporator sections of some turns are completely covered by liquid (Fig.~\ref{PHPviewer_woG}). Vigorous oscillations appear when the superheating of tube walls in the evaporator section become high enough to overcome the nucleation barrier.

\section{PHP stable operation at different orientations}\label{opReg}

\begin{figure*}[htb]
	\centering
	\subfigure[$-90^{\circ}$ orientation]
	{\includegraphics[width=5.5cm,clip]{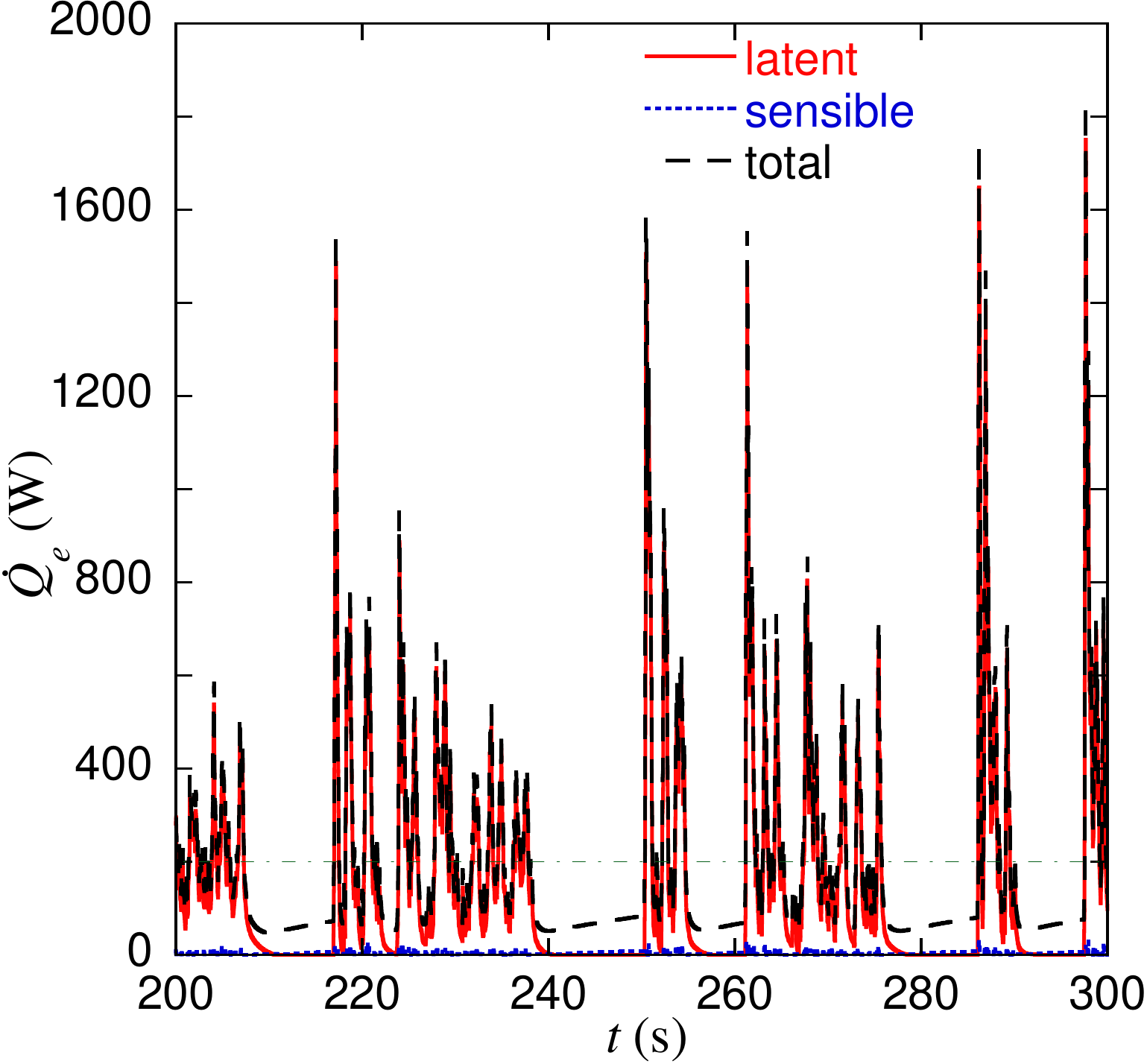}\label{Q_stable_aG}}
	\subfigure[$0^{\circ}$ orientation]
	{\includegraphics[width=5.5cm,clip]{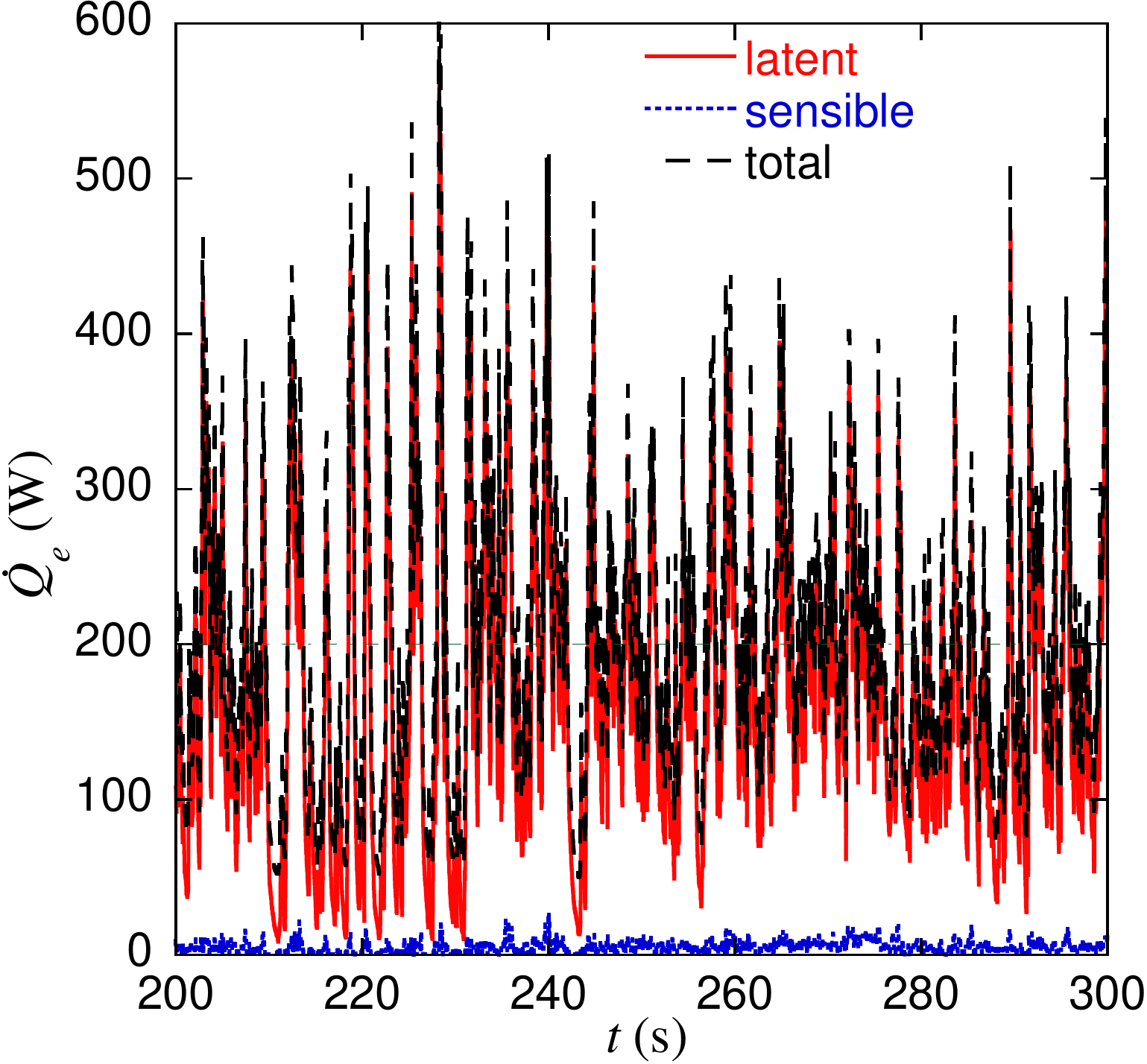}\label{Q_stable_woG}}
\subfigure[$90^{\circ}$ orientation]
	{\includegraphics[width=5.5cm,clip]{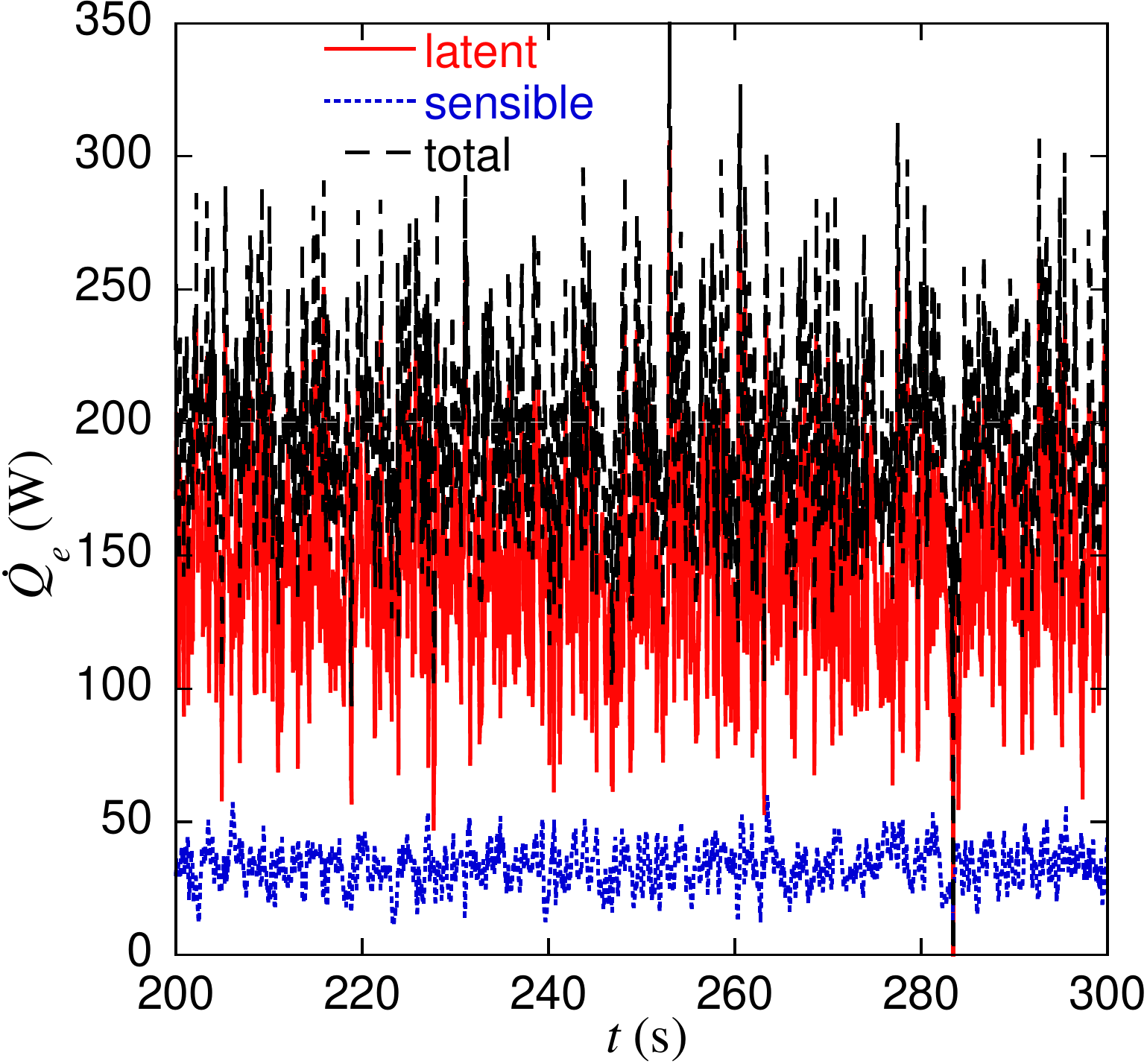}\label{Q_stable_wG}}
	\caption{Long-time evolution of the latent, sensible, and total heat exchange rates for $P_e=200$~W. The horizontal dash-dotted lines show $\langle\dot{Q}_e\rangle\simeq P_e$.}\label{Q_stable}
\end{figure*}

\cite{Karthikeyan14} described the regimes of stable operation which are observed experimentally in the PHP: continuous oscillations, intermittent or stopover regime and dryout. Our simulations also produce these regimes. Let us continue the description of the $-90^{\circ}$ case for $P_e=200$~W. The initial startup that occurs at $t\simeq 12$~s is short-lasting and stops at $t\simeq 14$~s. The liquid films deposited by the receding plug menisci slowly (with respect to the system dynamics during the initial oscillation burst) evaporate; $F_f$ remains to be nonzero for a much longer time than $F_l$, cf. Figs.~\ref{F_start_AG},\ref{Te_3cases}. The evaporator temperature keeps rising because $F_f$ is not large enough to provide efficient evaporator cooling. The system is thus again in the stopover configuration that lasts during the film drying. This second stopover is similar to the first stopover  discussed in sec.~\ref{StartUp}. At $t\simeq  21$~s, a new oscillation burst occurs and $T_e$ drops sharply again. Then another stopover follows. The stopovers and oscillation bursts (that correspond to the sharp $T_e$ drops in Fig.~\ref{Te_3cases}) alternate but their duration is irregular. This is a signature of the intermittency well known from the theory of dynamical chaos \citep{Ott2002}. Such a process finally leads to a stable regime (Fig.~\ref{TeAgLong}) where the temperature $T_e$ oscillates around an average value $\langle T_e\rangle$. This regime can be seen in Fig.~\ref{Q_stable_aG} where the heat exchanged with the evaporator section is analyzed.
The total power $\dot{Q}_e$ transferred from the evaporator can be divided into three distinctive parts: the latent heat $\dot{Q}_e^{lat}$ associated with the phase change at the surface of the liquid films and plug menisci, the heat $\dot{Q}_e^{sens}$ received by the fluid via the sensible heat transfer and the heat conducted via the tube. The latter is not represented separately in Figs.~\ref{Q_stable} but can be deduced from them by subtraction of two other contributions from $\dot{Q}_e$. The sensible heat exchange appears mostly due to the thermal interaction of the liquid plugs with the tube walls appearing during the oscillating motion inside the PHP. Another part of the sensible heat exchange (the vapor phase heat exchange with the walls appearing within the dry areas) that has been accounted for is much smaller.

The stopovers can be clearly seen in Fig.~\ref{Q_stable_aG} because during them both $\dot{Q}_e^{lat}$ and $\dot{Q}_e^{sens}$ drop sharply and the heat is transferred only through the tube walls; the wall contribution to $\dot{Q}_e$ slightly grows in time because during the stopovers $T_e$ grows (cf. Fig.~\ref{Te_3cases}). During the oscillation bursts, the instantaneous heat exchange is extremely strong and can be an order of value larger than $P_e$. The heat is transferred mainly as the latent heat; the sensible contribution is negligibly small because of the rare and brief residence of the liquid plugs in the evaporator. This can be seen from a very low average value of $F_l$, see Table \ref{table_contrib}. The predominance of the latent heat transfer (see Table \ref{table_contrib}) discussed earlier by \citet{JHT11} has been recently confirmed experimentally by \citet{Jo19} who measured experimentally its relative value situating between 0.66 and 0.74 for the $90^{\circ}$ case.
\begin{table}[htb]
	\centering
	\caption{Relative contributions of average latent and sensible heat transfer (normalized by $P_e$) and of coverage fractions of evaporator by different phases for different PHP orientations at $P_e=200$~W.}
	\begin{tabular}{|c|c|c|c|c|c|}
		\hline
       & $\langle\dot{Q}_e^{lat}\rangle$ &
$\langle\dot{Q}_e^{sens}\rangle$  & $\langle{F_f}\rangle$ &
$\langle{F_l}\rangle$ & $\langle{F_d}\rangle$ \\
        \hline
        $-90^{\circ}$& 0.73  & 0.02  & 0.14  & 0.01 & 0.85  \\
        \hline
        $0^{\circ}$&  0.80 & 0.02  & 0.24  & 0.04 & 0.72 \\
        \hline
        $90^{\circ}$& 0.73 & 0.17  & 0.42  & 0.37 & 0.21 \\
		\hline
	\end{tabular}
	\label{table_contrib}
\end{table}

A regime that can be characterized as intermediate between intermittent and continuous oscillations occurs for $P_e\geq 180$~W in the horizontal ($0^{\circ}$) orientation. It develops after an initial transient, where a long stopover typically occurs (15~s$<t<20$~s for $P_e=200$~W, cf. Fig.~\ref{Te_3cases}). Such a regime shows very short (one or several seconds) and rare complete stopovers. The stable regime is achieved after first 30~s (cf. the $0^{\circ}$ curve in Fig.~\ref{Te_3cases}; see the curve in Fig.~\ref{dtTe} below corresponding to $\Delta t=10^{-5}$~s for the long time evolution). In spite of almost continuous motion, the liquid plugs situate in the condenser and adiabatic sections most of the time. Their excursions to the evaporator are short-lasting so $\langle F_l\rangle$ is of the order of several percent (Fig.~\ref{F_start_woG}). This value is intermediate between the respective values for intermittent (Fig.~\ref{F_start_AG}) and continuous oscillations (Fig.~\ref{F_start_wG}) regimes.

When one performs the simulations for larger and larger power $P_e$, the  stopover periods become longer and longer until the oscillation bursts do not occur any more. Such a situation corresponds to the dryout.

\begin{figure}[htb]
	\centering
	\includegraphics[width=5.5cm,clip]{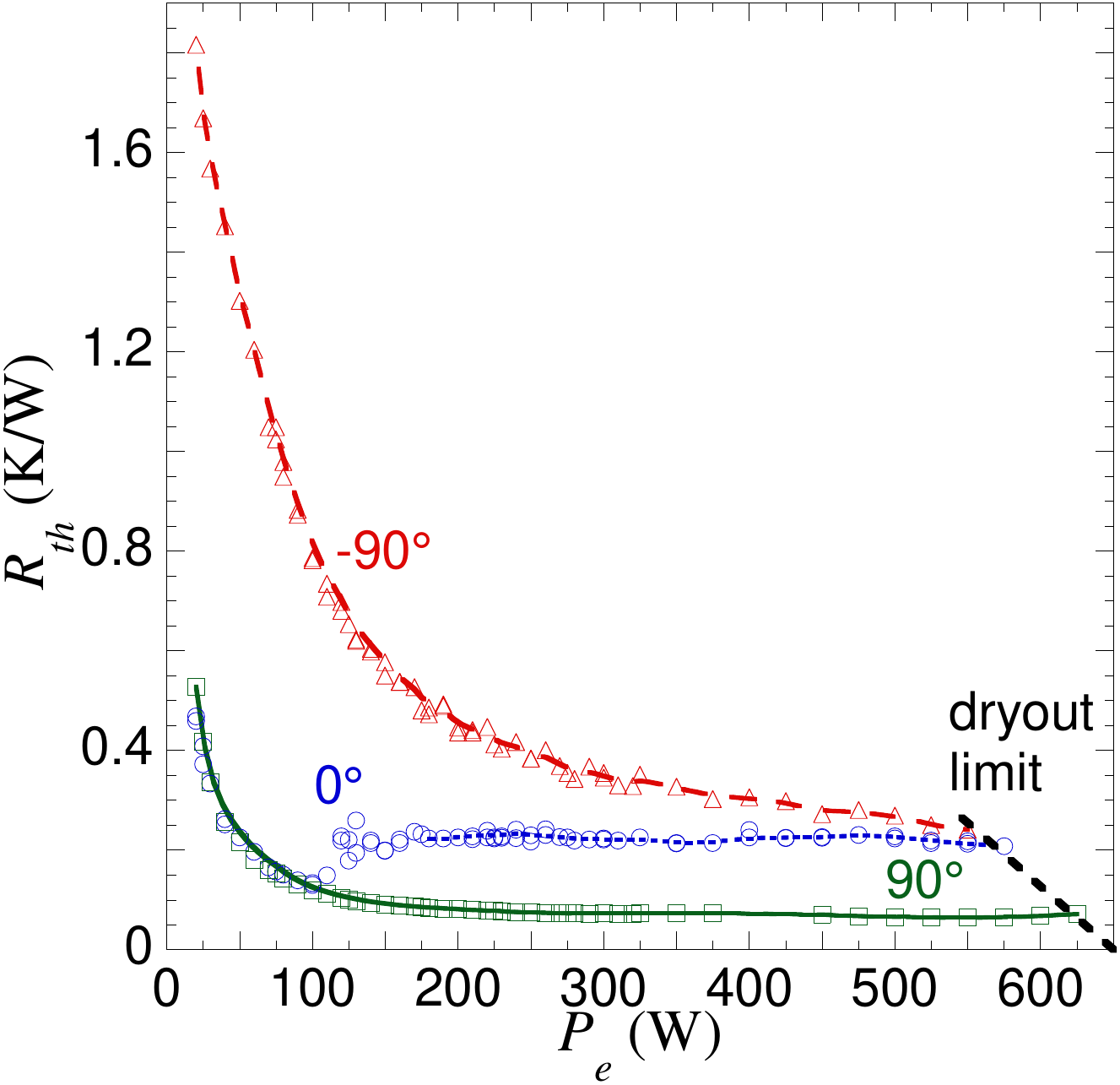}
	\caption{The PHP thermal resistance as a function of $P_e$ at different orientations. Triangles show the $-90^{\circ}$ data, circles, $0^{\circ}$ and squares, $90^{\circ}$. The lines are the eye guides.}\label{R_water}
\end{figure}

When $P_e<100$~W at the same horizontal orientation ($0^{\circ}$), the continuous oscillation regime is observed. The PHP then behaves exactly like in vertical favorable orientation ($90^{\circ}$). This can be seen in Fig.~\ref{R_water} where we present the dependence of the thermal resistance calculated as
\begin{equation}\label{Rth}
  R_{th}=\frac{\langle T_e\rangle-T_c}{P_e}
\end{equation}
on $P_e$. The data for $0^{\circ}$ and $90^{\circ}$ coincide in this region.

The continuous oscillation regime occurs almost in the whole $P_e$ range in the $90^{\circ}$ case. The evaporator cooling in continuous oscillation regime is more efficient than in the intermittent regime: $R_{th}$ values are lower in qualitative agreement with experimental results of \cite{Lips10}. This is explained by a much more frequent and durable residence of the liquid plugs (cf. Table \ref{table_contrib}) in the evaporator. The moving plugs deposit the liquid films that provide the efficient heat transfer. In addition, when the plugs situate within the evaporator, the bubble generation occurs inside the plugs. The bubbles escape quickly from the evaporator and their condensation occurs in the adiabatic section and condenser. Therefore, a permanent presence of the liquid plugs in the evaporator is a necessary condition for a good PHP performance. Accordingly to the frequent liquid plug presence in the evaporator, the sensible heat exchange (Fig.~\ref{Q_stable_wG}) can in this regime be comparable to the latent heat exchange.

The behavior of the thermal resistance with the heat load (Fig.~\ref{R_water}) is similar to the experimental observations of \cite{Lips10, Mangini15}. The curves for both vertical orientations are monotonously decreasing. The quantitative agreement is not bad either, at least the order of value is similar. This is satisfactory (the PHP structures are not the same). The curve for $0^{\circ}$ is non monotonous. As mentioned above, it nearly coincides with the $90^{\circ}$ curve at low $P_e$ (where the oscillations are continuous) and is thus decreasing too. At high $P_e$, $R_{th}$ is nearly constant and the oscillations are intermittent. There is a transition region at 100~W$<P_e<180$~W where the system is sensible to small change in simulation parameters. It should be recalled at this point that the system is chaotic, which means that a very small change in any parameter like initial condition or numerical precision (related e.g. to different processor models) results in different instantaneous values at a given $t$. However the simulation model is robust, which means that the average values of simulated variables are independent of these small changes. The situation is however different in the transition region where there is some data scatter shown in Fig.~\ref{R_water}.

\section{Simulation reliability}\label{code}

The heat and mass conservation in the system is important to be satisfied throughout the simulation. In any stable regime, the average heat power transferred from the evaporator should be equal to the amount received by the liquid: on average, there is no heat accumulation both in the tube wall and in the fluid. The sum $\dot Q_e$ of all the contributions (latent and sensible heat transfer through the fluid and the solid) should be equal to the input power $P_e$, see Fig.~\ref{Q_stable}. Since the oscillations are chaotic, a small error always appears in the summation. We verified that this error does not exceed at most one percent for the simulated time $t=5$ min.

\begin{figure}[htb]
	\centering
	\includegraphics[width=5.5cm,clip]{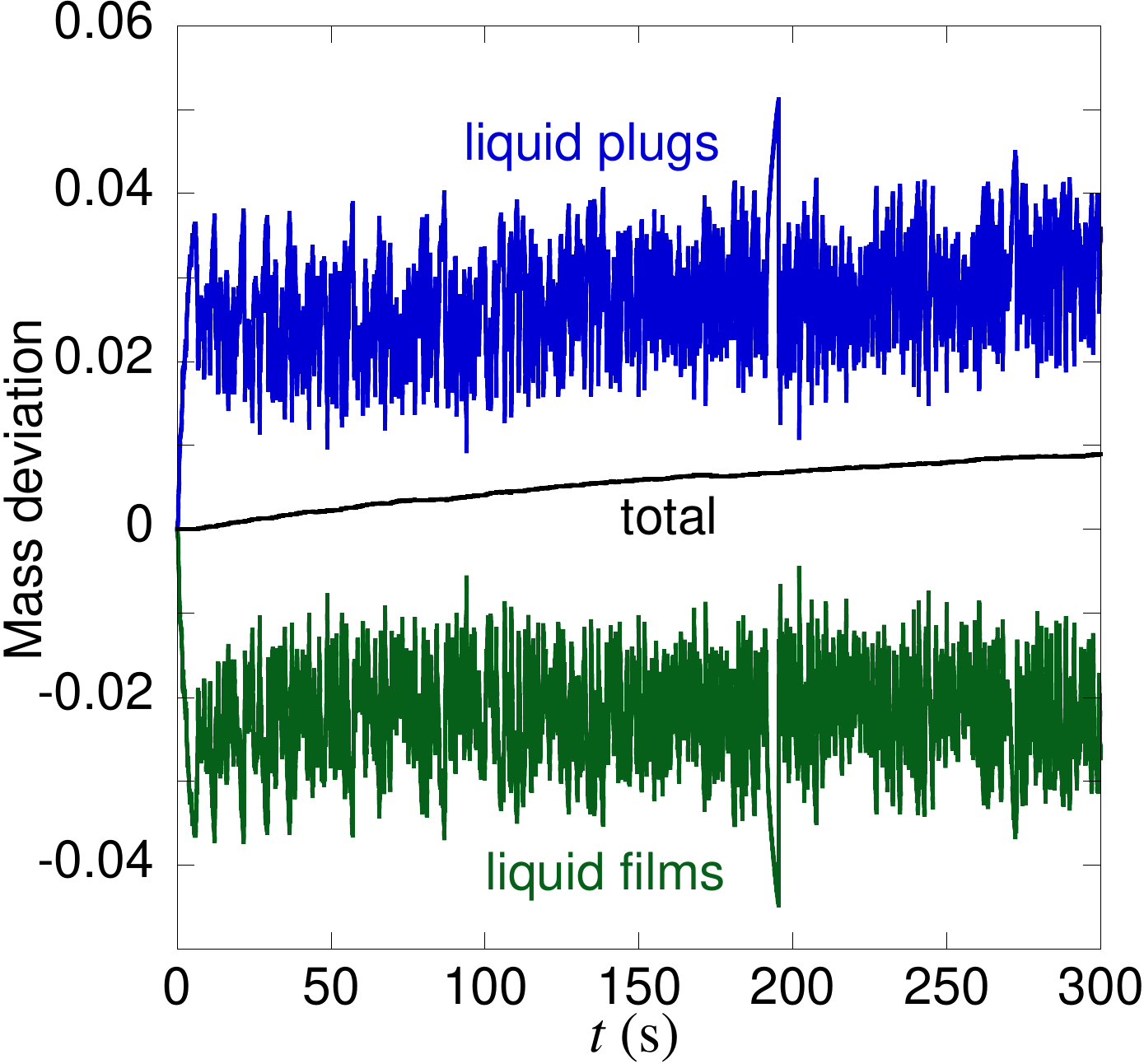}
	\caption{The relative mass deviation from the initial value for the liquid plugs, liquid films, and the total fluid mass during the simulation for $0^{\circ}$ and $P_e=400$~W.}\label{massConserv}
\end{figure}

The total fluid mass that consists of the mass of the liquid plugs, liquid films, and the vapor should ideally be conserved throughout the simulation. However, a small deviation always exists. The most evident reason is the overlap of the film edges and the menisci during their coalescence. Such an overlap appears because of the numerical error associated with the finiteness of the time step. The plug coalescence (i.e. bubble deletion) events are frequent, their number can be as large as $10^6$ during a run. The numerical algorithms used in CASCO were carefully worked out to minimize the mass deviation. In a typical continuous oscillation regime run, where the number of generated and deleted bubbles is of the order of $10^3$, the relative mass deviation is smaller than $10^{-4}$. A case of high thermal load where the regime is intermittent and the total number of the processed bubbles is about $10^5$ is shown in Fig.~\ref{massConserv}.

\begin{figure}[htb]
	\centering
	\includegraphics[width=5.5cm,clip]{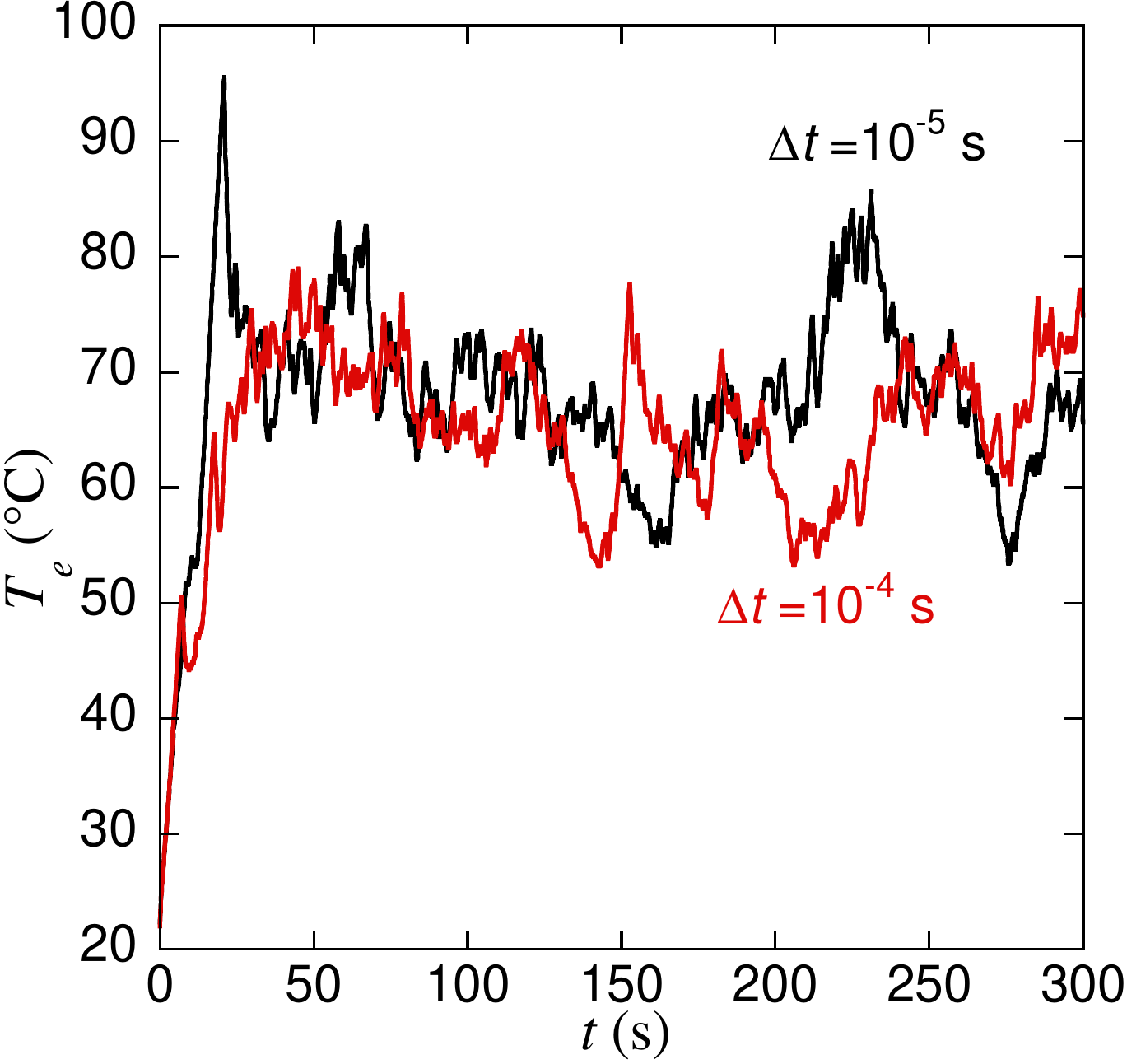}
	\caption{$T_e$ evolution calculated with two different time steps for $0^{\circ}$ and $P_e=200$~W.}\label{dtTe}
\end{figure}

Another important issue concerns the grid and time step independence of the results. It has been verified that the frequency of the liquid plug spatial remeshing applied during the simulation does not impact the results. The influence of the time step is shown in Fig.~\ref{dtTe} for the ``worst'' case of the intermittent regime that exhibits strong chaotic fluctuations. One can see that a change of the time step as large as ten times does not have a strong impact on the long-time $T_e$ average value after the start-up.

To give an idea about the CASCO efficiency, a typical simulation of 300 s of the PHP functioning takes about 8 hours on an Intel Core i7 based computer for $\Delta t=10^{-5}$~s and 1 hour for $\Delta t=10^{-4}$~s.
\section{Conclusions}

The 10-turn copper-water PHP with separately heated turns has been simulated for three different orientations with respect to gravity and for different power inputs.
The CASCO software is capable of long-time reliable simulations with a reasonably small required computer resources and good numerical stability.
In the vertical favorable orientation (evaporator below condenser) the continuous oscillations were observed for all heat loads. In the horizontal position, continuous oscillations occur until a threshold power, then a transition to the intermittent (stopover) regime was encountered. In the vertical unfavorable orientation (condenser below evaporator) the intermittent oscillations occur for all powers. The PHP dryout is observed for each regime when the heat load exceeds a threshold. The thermal resistance is obtained as a function of heat load. The obtained change of the thermal resistance with the heat load corresponds at least qualitatively to the experimentally obtained curves. The order of value of the resistance is similar to the experimental measurements. More detailed quantitative comparisons are necessary to tune the simulation parameters.

Our analysis shows that the difference in PHP performance at different orientations is linked to the difference in the phase distribution inside the PHP evaporator.

The contribution of the phase change to the overall heat exchange in the PHP is large in all the oscillation regimes. The relative contribution of the latent and sensible heat transfer rates controlled by the presence of the liquid plugs and liquid films inside the evaporator. The average fractions of evaporator sections covered by these phases are different at each PHP orientation. When PHP works efficiently (vertical favorable orientation and horizontal at small heat loads), the major part of the transferred heat is the latent heat. This agrees with recent experimental data. The sensible heat exchange through the fluid is usually larger than that through the tube walls. In the intermittent oscillation regime (vertical unfavorable and horizontal for large heat loads), the fluid sensible contribution is very low.

\section*{Acknowledgment}
The financial contribution of ESA within MAP INWIP is acknowledged. This work has been presented during the joint 19th International Heat Pipe Conference and 13th International Heat Pipe Symposium, Pisa, Italy (2018).

\bibliographystyle{spmpscinat}
\bibliography{PHP,Books}

\begin{thebibliography}{15}
\providecommand{\natexlab}[1]{#1}
\providecommand{\url}[1]{#1}
\providecommand{\urlprefix}{URL }
\expandafter\ifx\csname urlstyle\endcsname\relax
  \providecommand{\doi}[1]{DOI~\discretionary{}{}{}#1}\else
  \providecommand{\doi}{DOI~\discretionary{}{}{}\begingroup
  \urlstyle{rm}\Url}\fi

\bibitem[{Gu et~al.(2005)Gu, Kawaji, and Futamata}]{Gu05}
Gu, J., Kawaji, M., Futamata, R.: Microgravity performance of micro pulsating
  heat pipes.
\newblock Microgravity Sci. Technol. \textbf{16}(1), 181 -- 185 (2005).
\newblock \doi{10.1007/BF02945972}

\bibitem[{Holley and Faghri(2005)}]{Holley05}
Holley, B., Faghri, A.: Analysis of pulsating heat pipe with capillary wick and
  varying channel diameter.
\newblock Int. J. Heat Mass Transfer \textbf{48}(13), 2635 -- 2651 (2005).
\newblock \doi{10.1016/j.ijheatmasstransfer.2005.01.013}

\bibitem[{Jo et~al.(2019)Jo, Kim, and Kim}]{Jo19}
Jo, J., Kim, J., Kim, S.J.: Experimental investigations of heat transfer
  mechanisms of a pulsating heat pipe.
\newblock Energy Convers. Manage. \textbf{181}, 331 -- 341 (2019).
\newblock \doi{10.1016/j.enconman.2018.12.027}

\bibitem[{Karthikeyan et~al.(2014)Karthikeyan, Khandekar, Pillai, and
  Sharma}]{Karthikeyan14}
Karthikeyan, V., Khandekar, S., Pillai, B., Sharma, P.K.: Infrared thermography
  of a pulsating heat pipe: Flow regimes and multiple steady states.
\newblock Appl. Therm. Eng. \textbf{62}(2), 470 -- 480 (2014).
\newblock \doi{10.1016/j.applthermaleng.2013.09.041}

\bibitem[{Lips et~al.(2010)Lips, Bensalem, Bertin, Ayel, Romestant, and
  Bonjour}]{Lips10}
Lips, S., Bensalem, A., Bertin, Y., Ayel, V., Romestant, C., Bonjour, J.:
  Experimental evidences of distinct heat transfer regimes in pulsating heat
  pipes ({PHP}).
\newblock Appl. Therm. Eng. \textbf{30}(8-9), 900 -- 907 (2010).
\newblock \doi{10.1016/j.applthermaleng.2009.12.020}

\bibitem[{Mameli et~al.(2012)Mameli, Marengo, and Zinna}]{MameliMST12}
Mameli, M., Marengo, M., Zinna, S.: Numerical investigation of the effects of
  orientation and gravity in a closed loop pulsating heat pipe.
\newblock Microgravity Sci. Technol. \textbf{24}, 79 -- 92 (2012).
\newblock \doi{10.1007/s12217-011-9293-2}

\bibitem[{Mangini et~al.(2015)Mangini, Mameli, Georgoulas, Araneo, Filippeschi,
  and Marengo}]{Mangini15}
Mangini, D., Mameli, M., Georgoulas, A., Araneo, L., Filippeschi, S., Marengo,
  M.: A pulsating heat pipe for space applications: Ground and microgravity
  experiments.
\newblock Int. J. Therm. Sci. \textbf{95}, 53 -- 63 (2015).
\newblock \doi{10.1016/j.ijthermalsci.2015.04.001}

\bibitem[{Marengo and Nikolayev(2018)}]{EncycExp18}
Marengo, M., Nikolayev, V.: Pulsating heat pipes: Experimental analysis, design
  and applications.
\newblock In: Thome, J.R. (ed.) Encyclopedia of Two-Phase Heat Transfer and
  Flow {IV}, vol. 1: Modeling of Two-Phase Flows and Heat Transfer, pp. 1 --
  62. World Scientific (2018).
\newblock \doi{10.1142/9789813234406\_0001}

\bibitem[{Nekrashevych and Nikolayev(2017)}]{IarATE17}
Nekrashevych, I., Nikolayev, V.S.: Effect of tube heat conduction on the
  pulsating heat pipe start-up.
\newblock Appl. Therm. Eng. \textbf{117}, 24 -- 29 (2017).
\newblock \doi{10.1016/j.applthermaleng.2017.02.013}

\bibitem[{Nikolayev and Marengo(2018)}]{EncycSimu18}
Nikolayev, V., Marengo, M.: Pulsating heat pipes: Basics of functioning and
  numerical modeling.
\newblock In: Thome, J.R. (ed.) Encyclopedia of Two-Phase Heat Transfer and
  Flow {IV}, vol. 1: Modeling of Two-Phase Flows and Heat Transfer, pp. 63 --
  139. World Scientific (2018).
\newblock \doi{10.1142/9789813234406\_0002}

\bibitem[{Nikolayev(2011)}]{JHT11}
Nikolayev, V.S.: A dynamic film model of the pulsating heat pipe.
\newblock J. Heat Transfer \textbf{133}(8), 081504 (2011).
\newblock \doi{10.1115/1.4003759}

\bibitem[{Nikolayev(2013)}]{IJHMT13}
Nikolayev, V.S.: Oscillatory instability of the gas-liquid meniscus in a
  capillary under the imposed temperature difference.
\newblock Int. J. Heat Mass Transfer \textbf{64}, 313 -- 321 (2013).
\newblock \doi{10.1016/j.ijheatmasstransfer.2013.04.043}

\bibitem[{Nikolayev and Nekrashevych(2018)}]{IHPC18VN}
Nikolayev, V.S., Nekrashevych, I.: Vapor thermodynamics and fluid merit for
  pulsating heat pipe.
\newblock In: Proc. 19th Int. Heat Pipe Conf. and 13th Int. Heat Pipe Symp.
  Pisa, Italy (2018)

\bibitem[{Ott(2002)}]{Ott2002}
Ott, E.: Chaos in Dynamical Systems.
\newblock 2 edn. Cambridge University Press (2002).
\newblock \doi{10.1017/CBO9780511803260}

\bibitem[{Shafii et~al.(2001)Shafii, Faghri, and Zhang}]{shafii1}
Shafii, M.B., Faghri, A., Zhang, Y.: Thermal modeling of unlooped and looped
  pulsating heat pipes.
\newblock J. Heat Transfer \textbf{123}(6), 1159 -- 1172 (2001).
\newblock \doi{10.1115/1.1409266}

\end{thebibliography}

\end{document}